\documentclass[aps,pre,twocolumn,showpacs]{revtex4}
\usepackage{epsfig}
\usepackage{times}
\usepackage{amsmath}
\begin{document}

\title{Restricted connections among distinguished players support cooperation}

\author{Matja{\v z} Perc,$^1$ Attila Szolnoki,$^2$ and Gy\"{o}rgy Szab\'{o}$^2$}

\affiliation
{$^1$Department of Physics, Faculty of Natural Sciences and Mathematics, University of \\ Maribor, Koro{\v s}ka cesta 160, SI-2000 Maribor, Slovenia \\
$^2$Research Institute for Technical Physics and Materials Science, P.O. Box 49, H-1525 Budapest, Hungary}

\begin{abstract}
We study the evolution of cooperation within the spatial prisoner's dilemma game on a square lattice where a fraction of players $\mu$ can spread their strategy more easily than the rest due to a predetermined larger teaching capability. In addition, players characterized with the larger teaching capability are allowed to temporarily link with distant opponents of the same kind with probability $p$, thus introducing shortcut connections among the distinguished. We show that these additional temporary connections are able to sustain cooperation throughout the whole range of the temptation to defect. Remarkably, we observe that as the temptation to defect increases the optimal $\mu$ decreases, and moreover, only minute values of $p$ warrant the best promotion of cooperation. Our study thus indicates that influential individuals must be few and sparsely connected in order for cooperation to thrive in a defection prone environment.
\end{abstract}


\maketitle

\section{Introduction}
Sustenance of cooperation within groups of selfish individuals is a challenge faced by scientists across fields of research as different as sociology, economics and biology \cite{axelrod_84}. The essence of the problem lies in the fact that cooperation implies working for mutual interests or the common good of society on the expense of individual prosperity. The additional costs of cooperation can be avoided by choosing defection, and accordingly, the cheating behavior of defectors spreads if the evolutionary process is governed by the imitation of more successful strategies. However, as the defectors become dominant the whole society suffers because nobody remains that would contribute to the overall welfare, hence the dilemma. A commonly adopted framework for addressing the issue is the evolutionary game theory \cite{weibull_95, hofbauer_98, gintis_00, nowak_06}, and the prisoner's dilemma game in particular, which in its well-mixed version reflects exactly the described plundering of defectors and the consequent extinction of cooperators.

Although mechanisms such as kin selection, direct and indirect reciprocity or voluntary participation are largely successful in preventing the defectors to reign \cite{nowak_s06}, the seminal observation promoting the survival of the cooperative trait arguably came in the form of spatial games \cite{nowak_n92b, lindgren_pd94}, where the participating players no longer abide to the principles of well-mixed dynamics, but instead, cooperators are able to survive via clustering that protects them mutually against the exploitation by invading defectors (for a recent review, see \cite{szabo_pr07}). Another important development that facilitated the understanding of the evolution of cooperation came in the form of replacing the initially proposed regular interaction scheme with more complex topologies \cite{abramson_pre01, ebel_pre02, holme_pre03,  masuda_pla03, wu_pre05, lieberman_n05, tomassini_pre06, vukov_pre08, wang_pre06, ren_pre07, chen_pre08, luthi_pa08}, whereby in particular the scale-free network has been identified as an excellent host for cooperative individuals \cite{santos_prl05, santos_prslb06}, warranting the best protection against the defectors. Since the strong heterogeneity of the degree distribution on scale-free networks was identified as the main driving force behind the flourishing cooperative state \cite{santos_jeb06, gomez-gardenes_prl07, masuda_prsb07, poncela_njp07, szolnoki_pa08}, some alternative sources of inhomogeneity were already investigated as potential promoters of cooperation with noticeable success. Recent examples of such approaches include the introduction of preferential selection \cite{wu_pre06}, asymmetry of connections \cite{kim_pre02}, different teaching capabilities \cite{szolnoki_epl07}, heterogeneous influences \cite{wu_cpl06}, or social diversity \cite{perc_pre08}. Arguably, the differences between participating players, either in terms of their degree, teaching capability or social rank, are easily justified from the viewpoint of real life societies, as the latter are in general soaked with members of different status having unlike opportunities to become influential in the future. This may be especially obvious by humans, but by no means difficult to observe in animal societies as well.

At present, our goal is to extend the scope of beneficial influences of heterogeneities on the evolution of cooperation by considering a spatial prisoner's dilemma game where players differ not only in their teaching capabilities, but in addition, the distinguished players posses the ability to temporarily connect with distant individuals of the same rank and try to overtake them. We show that this fairly simple additional extension may provide an unprecedented boost for cooperators that can be compared only to the facilitative effect warranted by the scale-free topology if using absolute payoffs. Indeed, for an optimal fraction of distinguished teachers $\mu$ and probability to temporarily link them during the evolutionary process $p$, the defectors remain outnumbered throughout the whole span of the temptation of defect $b$. Although intuitively one might expect that larger $b$ would require increasing numbers of strongly connected leaders to sustain cooperation, we reveal that in fact the optimal $\mu$ decreases continuously as $b$ increases, and also, the interconnectedness of the distinguished determined via $p$ has to remain very weak in order for cooperation to thrive best. We study the mechanism underlying the reported promotion of cooperation by calculating temporal courses of cooperator densities separately for the distinguished players and for their interacting nearest neighbors. In addition, we discuss our findings in view of recent results obtained on scale-free networks under assortative and disassortative mixing \cite{rong_pre07}, and emphasize that special complex topologies may not be a necessary ingredient of a flourishing cooperative society.

The remainder of this paper is organized as follows. In the next section we describe the employed spatial prisoner's dilemma game and other details of the evolutionary process. Section III is devoted to the presentation of results, whereas in the last section we summarize and discuss their implications.

\section{Game definition and setup}
As noted, we use the spatial prisoner's dilemma game for the purpose of this study, which in accordance with the parametrization suggested by Nowak and May \cite{nowak_n92b} is characterized by the temptation $T = b$, reward $R = 1$, and both punishment $P$ as well as the suckers payoff $S$ equaling $0$, whereby $1 < b \leq 2$ ensures a proper payoff ranking. The game is staged on a regular $L \times L$ square grid with nearest neighbor interactions and periodic boundary conditions, whereon initially each player on site $x$ is designated either as a cooperator ($s_x = C$) or defector ($D$) with equal probability. Forward iteration is performed in accordance with the Monte Carlo simulation procedure comprising the following elementary steps. First, a randomly selected player $x$ acquires its payoff $P_x$ by playing the game with its four nearest neighbors. Next, one randomly chosen neighbor, denoted by $y$, also acquires its payoff $P_y$ by playing the game with its nearest neighbors. Last, player $x$ tries to enforce its strategy $s_x$ on player $y$ in accordance with the probability
\begin{equation}
W(s_y \rightarrow s_x)=w_x \frac{1}{1+\exp[(P_y-P_x)/K]},
\label{eq:prob}
\end{equation}
where $K$ denotes the amplitude of noise and $w_x$ characterizes the teaching capability of player $x$. The parameter $w_x$ is assigned to each player at the beginning of the game and remains fixed during the evolutionary process. In particular, among all $L^2$ players, and irrespective of their initial strategies, a fraction $\mu$ is chosen randomly and designated as having $w_x  = 1$ whereas the remaining $1 - \mu$ are assigned $w_x = 0.01$. Players within the former group are the so-called distinguished players (or teachers) that are characterized with the larger teaching capability, and according to Eq.~(\ref{eq:prob}), are much more likely to reproduce than individuals pertaining to the less influential (or blocked) group. Noteworthy, a similar setup has been considered in \cite{szolnoki_epl07} where the parameter $\nu$ determined the fraction of blocked players. Thus, a direct link to the present study can be established by acknowledging that $\mu = 1 - \nu$. Moreover, the phase diagram of the prisoner's dilemma game on a square lattice for a given $\nu$ presented in \cite{szolnoki_epl07} reveals that the cooperation facilitative effect of distinguished players becomes better pronounced at high $K$. We will therefore use $K=2$ throughout this work, except in Fig.~\ref{fig:fig4} where absorbing cooperative states would prohibit useful comparisons of results obtained at different $b$, in which case $K=0.4$ will be used. Worthy of notice is also that the two limiting cases $\mu = 0$ and $\mu = 1$ result in homogeneous teaching capability assigned to all involved and are thus equal, only that in the former case the evolutionary process is slower.

Further upgrading the model, we introduce the possibility of direct information transfer between distinguished players that are characterized by $w_x  = 1$. In particular, a teacher from within the group of distinguished players may choose with probability $p$, instead of a nearest neighbor with probability $1-p$, a distant randomly selected other teacher to be the target for strategy transfer. It is important to note that thereby only the strategy transfer is allowed between the two distant teachers, yet both still collect their payoffs by playing the prisoner's dilemma game with their four nearest neighbors, as described above. Thus, $p$ is simply the probability of temporarily interconnecting two distant distinguished players during an elementary part of the Monte Carlo step, whereas the remaining steps of the evolutionary process are left the same. This directly implies that our findings are independent of payoff normalization as the latter simply scales $K$ but does not introduce qualitatively different results. It is also worth noting that permanent connections between members of the group having $w_x  = 1$ result in similar behavior as will be reported below, yet presently we wanted to avoid effects that might be caused by differences in the degree of permanently linked distant players.

Monte Carlo results presented below were obtained on populations comprising $300 \times 300$ to $800 \times 800$ individuals, whereby the fraction of cooperators $\rho_C$ was determined within $2\cdot10^5$ to $2\cdot10^6$ full MC steps (MCS) after the transients were discarded. It is worth noting that the above introduced dynamical rule can be interpreted as a Markov chain with two absorbing states ($C$ or $D$), where thus the observed mixed states can be referred to as being stationary only for infinitely large system sizes, whereas for finite systems it is more appropriate to speak of quasi-stationary states or rather fixation probabilities of the two strategies and average times needed to reach the truly stationary absorbing states. Throughout the next section parameters $\mu$ and $p$ will be devoted the most attention to as they are crucial in determining the density and interconnectedness of distinguished players on the grid.

\section{Results}

\begin{figure}
\centerline{\epsfig{file=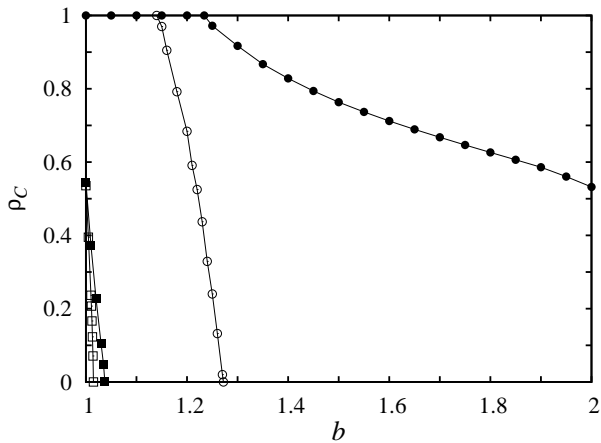,width=8cm}}
\caption{\label{fig:fig1}Fraction of cooperators $\rho_C$ in dependence on the temptation to defect $b$ obtained by setting: $\mu = 1$ and $p = 0$ (open squares), $\mu = 1$ and $p = 0.03$ (closed squares), $\mu = 0.12$ and $p = 0$ (open circles), $\mu = 0.12$ and $p = 0.03$ (closed circles). Only the joint adjustment of $\mu$ and $p$ warrants supreme promotion of cooperation. Lines are just guides for the eye.}
\end{figure}

We start by comparing results obtained with the presently introduced evolutionary model and its simplified versions to stress the joint relevance of the two main parameters $\mu$ and $p$. Figure~\ref{fig:fig1} features $\rho_C$ in dependence on $b$ for four different cases. The fastest decaying $\rho_C$ is obtained via the classical spatial prisoner's dilemma game where all players are characterized by $w_x = 1$ ($\mu = 1$) and temporary shortcut links among distant players are disabled ($p = 0$). Slightly better results in terms of cooperation sustainability are obtained if the latter condition is relaxed by setting $p = 0.03$, thus allowing rare temporary deviation from the nearest neighbor structure (here $\mu$ is still 1). Strikingly better results, on the other hand, are obtained if instead of $p$ the parameter $\mu$ is varied. Open circles show results obtained for $p = 0$ and $\mu = 0.12$, whereby the model with $p = 0$ has been studied extensively in \cite{szolnoki_epl07} and the interested reader may seek additional information on the effects of different $\mu$ there. Clearly the best environment for cooperators, however, is warranted when both $\mu$ and $p$ are adjusted. Indeed, by setting $\mu = 0.12$ and $p = 0.03$ we achieve that $\rho_C > 0.5$ throughout the whole range of $b$, as depicted by closed circles in Fig.~\ref{fig:fig1}. Thus, the joint impact of appropriate $\mu$ and $p$ strongly favors the sustainability of cooperation to the extend comparable only to effects observed previously on scale-free networks if using absolute payoffs \cite{santos_prl05}.

\begin{figure}
\centerline{\epsfig{file=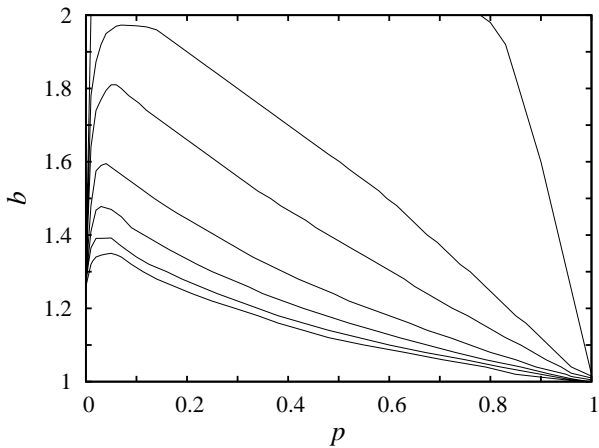,width=8cm}}
\caption{\label{fig:fig2}Contour line plot of $\rho_C$ in dependence on $p$ and $b$, obtained for $\mu = 0.3$. Lines mark $\rho_C$ equalling $1$, $0.8$, $0.6$, $0.4$, $0.2$, $0.1$ and $0$ from bottom to top. There exist an optimal value of $p \approx 0.05$ that promotes cooperation best.}
\end{figure}

In order to examine the impact of the newly introduced parameter $p$ more precisely, we present in Fig.~\ref{fig:fig2} a contour line plot showing the dependence of $\rho_C$ on $p$ and $b$ at a fixed value of $\mu = 0.3$. It can be inferred at glance that there exists an optimal value of $p$ warranting the best promotion of cooperation, which by the selected value of $\mu$ equals $p \approx 0.05$. Most importantly however, it is crucial to note the immense improvement in $\rho_C$ that is brought about by the addition of rare temporary long-range connections among distinguished players. In particular, while for $p = 0$ cooperators go extinct at $b = 1.3$, they prevail up to $b = 2.0$ in case $p$ is fine-tuned. In addition, the span of complete dominance is markedly enhanced as well. We argue that the role of distinguished players in the small $\mu$ region is similar to the role of players occupying the hubs of a scale-free network, whereby the temporary long-range connections enable them to enforce cooperative behavior not just to their permanently linked nearest neighbors but to distant players as well, who then in turn spread the cooperative trait further to their nearest neighbors, and so on, thus resulting in an optimal environment for the survival of cooperators even if temptations to defect are large.

\begin{figure}
\centerline{\epsfig{file=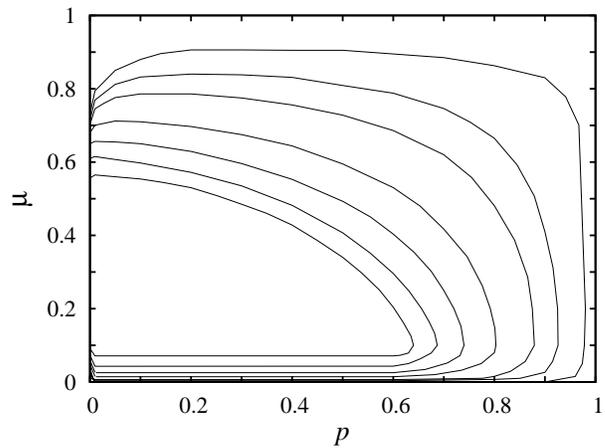,width=8cm}}
\caption{\label{fig:fig3}Contour line plot of $\rho_C$ in dependence on $p$ and $\mu$, obtained for $b = 1.1$. Lines mark $\rho_C$ equalling $1$, $0.8$, $0.6$, $0.4$, $0.2$, $0.1$ and $0$ from left to right. Note the double resonance in cooperation that peaks at small $p$ and $\mu$.}
\end{figure}

\begin{figure}
\centerline{\epsfig{file=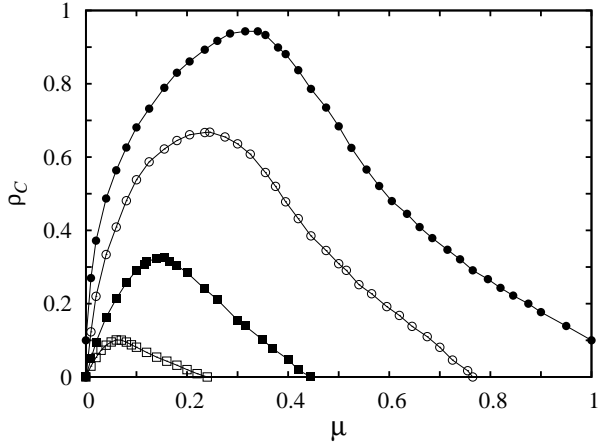,width=8cm}}
\caption{\label{fig:fig4}Fraction of cooperators $\rho_C$ in dependence on $\mu$ obtained by setting $p = 0.4$ and $K = 0.4$ (note that the lower value of $K$ was chosen solely to prohibit extensive absorbing cooperative states, hence enabling more accurate comparisons). Results are depicted for different temptations to defect: $b = 1.1$ (closed circles), $b = 1.2$ (open circles), $b = 1.5$ (closed squares) and $b = 2$ (open squares). The optimal value of $\mu$ decreases continuously as $b$ increases. Lines are just guides for the eye.}
\end{figure}

Furthermore, it is instructive to examine how $\rho_C$ varies in dependence on $p$ and $\mu$. Figure~\ref{fig:fig3} reveals that a double resonance in cooperation, induced by variations of $p$ and $\mu$, characterizes this dependence, thus suggesting that a fine tuning of both parameters is necessary for designing the optimal environment for cooperative behavior. In order to examine the resonant-like outlay of $\rho_C$ in dependence on $\mu$ more precisely, Fig.~\ref{fig:fig4} features results obtained for different values of $b$ and a fixed value of $p$. Notably, the optimal $\mu$ decreases continuously as $b$ increases, yet its careful adjustment may still propel cooperation away from extinction.

\begin{figure}
\centerline{\epsfig{file=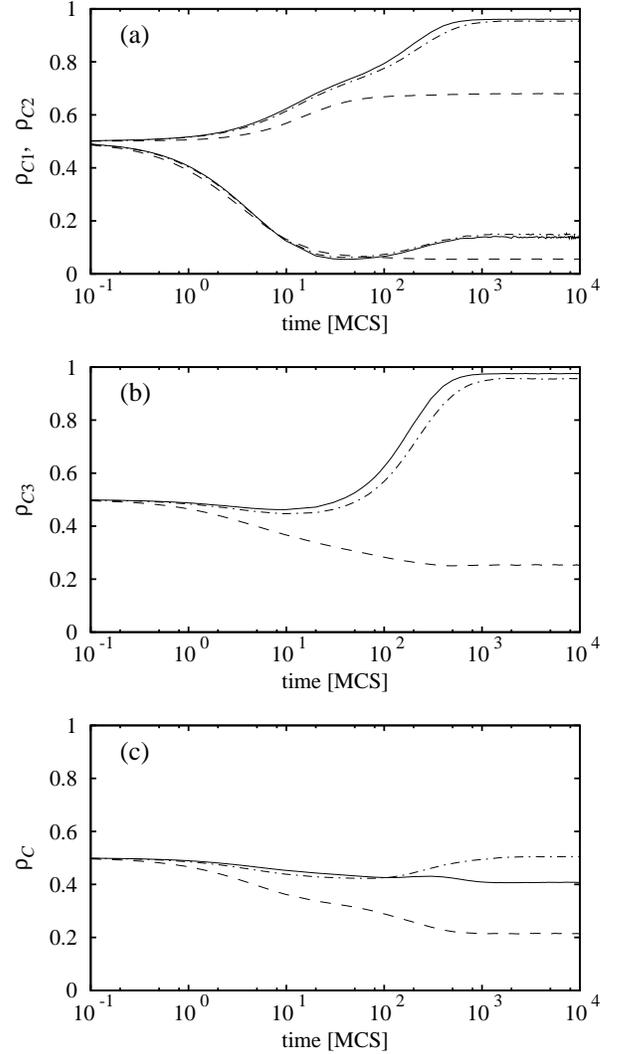,width=8cm}}
\caption{\label{fig:fig5}Time evolution of $\rho_C$, $\rho_{C1}$ [lower three curves in panel (a)], $\rho_{C2}$ [upper three curves in panel (a)] and $\rho_{C3}$ for $\mu=0.07$ (solid line), $\mu=0.1$ (dash-dotted line) and $\mu=0.25$ (dashed line), obtained by setting $p = 0.03$ and $b = 2$. See main text for definitions of $\rho_{C1}$, $\rho_{C2}$ and $\rho_{C3}$.}
\end{figure}

To understand the impact of different values of $\mu$, we recall the feedback mechanism resulting in widely enhanced cooperation within the model where players occupied a scale-free network \cite{santos_prl05}. There hubs can dominate over their neighborhoods because a larger degree directly results in a larger payoff. Hence the subordinate neighbors will imitate the hub's strategy, eventually producing homogeneous clouds of a given strategy around each hub. During this process the nature of the defecting (cooperating) strategy weakens (strengthens) the governing hub, which in turn leads to an easy victory of a cooperator hub when faced with a defector hub and thus to the widespread dissemination of the cooperative trait. In the present model a similar feedback mechanism is at work, but only when the distinguished players are sparse enough not to have their neighborhoods affected by other, potentially defecting, influential players; particularly when they are not directly linked with one another and they don't share mutual neighbors. Therefore we argue that a primary estimate for this condition to be fulfilled is $\mu < \theta_s = 0.1869(1)$, whereby $\theta_s$ is the jamming coverage of particles during a random sequential adsorption \cite{flory_jacs39, widow_jcp66} when nearest- and next-nearest neighbor interactions are excluded on a square lattice \cite{dickman_jcp91}. We find that for our model the more accurate value of the jamming coverage for the case when distinguished players don't share mutual neighbors is $\mu_c = 0.13965(5)$, thus validating the initial estimate via $\theta_s$. In this low $\mu < \mu_c$ region the previously described feedback mechanism can work because the distinguished players can impose their strategies on the neighbors without being disturbed. To illustrate this process we monitored how the density of cooperators evolves within different subgroups of the whole population. In particular, besides the density of cooperators among all $L^2$ players denoted by $\rho_C$, we also measure the density of cooperators among all the nearest neighbors of distinguished defectors (cooperators), which we denote by $\rho_{C1}$ ($\rho_{C2}$), and the density of cooperators among the distinguished players, which we denote by $\rho_{C3}$. Obtained results are presented in Fig.~\ref{fig:fig5} for three different values of $\mu$, whereby $\mu=0.25$ is higher than the critical $\mu_c$ value, $\mu=0.1$ is the optimal and $\mu=0.07$ the below-optimal value at $b=2$. A two-stage process can be inferred by following the time courses of the four calculated cooperator densities, which gives insights into the mechanism underlying the promotion of cooperation.

First, slightly prior to reaching the $100$ MCS, defecting (cooperating) distinguished players spread their strategy successfully among their neighbors, as evidenced by the local minima in Fig.~\ref{fig:fig5}(a). As soon as the minimum in $\rho_{C1}$ is reached, the second part of the two-stage process begins, which involves turning the defecting distinguished players into distinguished cooperators. In particular, as defectors occupy virtually the whole neighborhood of a distinguished defector, the latter becomes extremely weak because there is no one left to exploit. Thus, as soon as an influential cooperator receives the opportunity to overtake the weakened defector via a temporary long-range connection the latter is defeated, and the newly seeded cooperator starts spreading. Note that the described two-stage process, including temporary minimum of $\rho_{C1}$, cannot be observed at high values of $\mu$ exceeding $\mu_c$. There the distinguished cooperators cannot be successful because their neighbors may be exploited by other distinguished defectors. This feature is demonstrated in Fig.~\ref{fig:fig5}(b), where the cooperator density amongst the distinguished $\rho_{C3}$ remains low if $\mu=0.25$, but raises markedly for $\mu < \mu_c$. However, while very low values of $\mu$ (below the optimal) enable distinguished defectors to convert virtually all their neighbors to defectors, and thus make the negative feedback effect destined to work, at the end only a few distinguished cooperators resulting from the two-stage process cannot sustain an overall high level of cooperation. Hence, an optimal $\mu$ exists which still initializes the feedback mechanism, but subsequently warrants also that the density of distinguished cooperators is high enough to sustain the highest level of cooperation within the whole population, as evidenced in Fig.~\ref{fig:fig5}(c).

We emphasize that the above described mechanism can work even for large $b$, where substantial portions of non-distinguished players are controlled by the spatial evolutionary rule strongly favoring defection. In this situation the cooperative behavior can nevertheless prevail due to small cooperative colonies that can form around isolated distinguished players and then survive for very long times. Naturally, this mechanism of maintaining small cooperative islands is present also at lower values of $b$, but there the defecting strategy is not so beneficial among non-distinguished players, and hence the relative contribution of such small colonies to the overall cooperation level is moderate. This is why the impact on the evolution of cooperation is most evident at high values of $b$.

\begin{figure}
\centerline{\epsfig{file=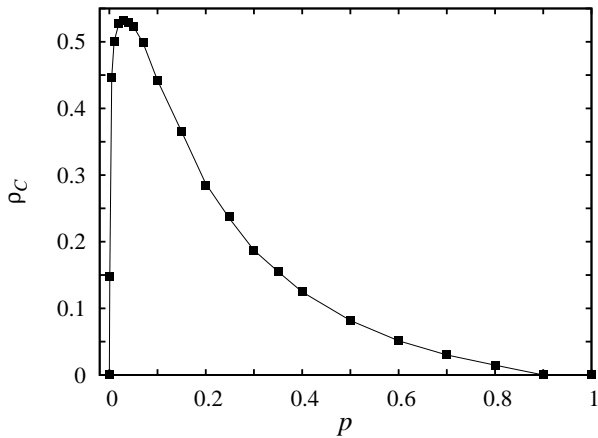,width=8cm}}
\caption{\label{fig:fig6}Fraction of cooperators $\rho_C$ in dependence on $p$, obtained by setting $\mu = 0.12$ and $b = 2$. Remarkably, the optimal $p$ is very small ($\simeq 0.03$). The line is just a guide for the eye.}
\end{figure}

To emphasize the necessity of a weak temporary interconnectedness of distinguished players, Fig.~\ref{fig:fig6} shows $\rho_C$ in dependence on $p$ for $\mu = 0.12$. The existence of an optimal $p$ can be observed clearly, and indeed, as little as $p = 0.03$ yields the maximum value of $\rho_C$. For lower values of $p$ the isolated influential cooperators can be eliminated by stochastic events long before they are able to pass their strategy to defecting distinguished players, while the increase of $p$ drives the system towards the mean-field type behavior favoring defection over cooperation.

The latter observation can be corroborated by some conceptually similar findings presented recently by Rong et al. \cite{rong_pre07} who studied the role of different degree mixing patterns on scale-free networks. There the assortative mixing, tending to interconnect the hubs, was also found to diminish the level of cooperation, whereas the disassortative mixing, promoting the isolation of hubs, further enhanced the cooperative trait for very large $b$ but decreased the density of cooperators in noisy environments for moderate temptations to defect. Presently, we show that the special scale-free topology is not a necessary ingredient for this type of cooperation facilitation, as in our case the uncorrelated rare random links may also provide the most favorable frequency of connections between distinguished players to optimally promote cooperative behavior.

\section{Summary}
In sum, we show that the additional introduction of temporary long-range connections among distinguished players in a heterogeneous population comprising two different types of individuals warrants a substantial promotion of cooperation within the evolutionary prisoner's dilemma game on a regular lattice. The joint effect of heterogeneity and temporary long-range connections is capable to maintain cooperation within the whole range of temptations to defect ($1 < b \leq 2$) that are usually considered for iterated prisoner's dilemma games. Noteworthy, our approach bears some similarity with game theoretical models entailing the co-evolution of strategy and network structure \cite{zimmermann_pre04, zimmermann_pre05, pacheco_jtb06, pacheco_prl06}, albeit presently the evolution of the interaction network in terms of temporary shortcuts among the distinguished players is completely random. Moreover, we reveal that environments which are strongly prone to defection require modest densities of influential players, which in addition, must not be strongly interrelated with one another. Importantly though, in the complete absence of interconnectedness these potential sources of cooperative behavior are unable to enforce the strategy on more than just their immediate neighbors, and hence a positive yet small value of $p$ provides just the missing virtue that enables the influential players to fully exploit their potentials. Our study thus indicates that, while in a modestly corrupted society characterized by small $b$ influential players may be many and well connected, this proves fatal in strongly defection prone environments. The latter, on the other hand, require isolated and weakly connected sources of cooperative behavior, which on one hand give defectors enough space to completely weaken their neighborhoods, and on the other, are frequent and interconnected enough to overtake these sites after the negative feedback has kicked in.

\begin{acknowledgments}
The authors acknowledge support from the Slovenian Research Agency (grant Z1-9629) and the Hungarian National Research Fund (grant K-73449).
\end{acknowledgments}


\begin{thebibliography}{38}
\expandafter\ifx\csname natexlab\endcsname\relax\def\natexlab#1{#1}\fi
\expandafter\ifx\csname bibnamefont\endcsname\relax
  \def\bibnamefont#1{#1}\fi
\expandafter\ifx\csname bibfnamefont\endcsname\relax
  \def\bibfnamefont#1{#1}\fi
\expandafter\ifx\csname citenamefont\endcsname\relax
  \def\citenamefont#1{#1}\fi
\expandafter\ifx\csname url\endcsname\relax
  \def\url#1{\texttt{#1}}\fi
\expandafter\ifx\csname urlprefix\endcsname\relax\def\urlprefix{URL }\fi
\providecommand{\bibinfo}[2]{#2}
\providecommand{\eprint}[2][]{\url{#2}}

\bibitem[{\citenamefont{Axelrod}(1984)}]{axelrod_84}
\bibinfo{author}{\bibfnamefont{R.}~\bibnamefont{Axelrod}},
  \emph{\bibinfo{title}{The Evolution of Cooperation}}
  (\bibinfo{publisher}{Basic Books}, \bibinfo{address}{New York},
  \bibinfo{year}{1984}).

\bibitem[{\citenamefont{Weibull}(1995)}]{weibull_95}
\bibinfo{author}{\bibfnamefont{J.~W.} \bibnamefont{Weibull}},
  \emph{\bibinfo{title}{Evolutionary Game Theory}} (\bibinfo{publisher}{MIT
  Press}, \bibinfo{address}{Cambridge, MA}, \bibinfo{year}{1995}).

\bibitem[{\citenamefont{Hofbauer and Sigmund}(1998)}]{hofbauer_98}
\bibinfo{author}{\bibfnamefont{J.}~\bibnamefont{Hofbauer}} \bibnamefont{and}
  \bibinfo{author}{\bibfnamefont{K.}~\bibnamefont{Sigmund}},
  \emph{\bibinfo{title}{Evolutionary Games and Population Dynamics}}
  (\bibinfo{publisher}{Cambridge University Press},
  \bibinfo{address}{Cambridge}, \bibinfo{year}{1998}).

\bibitem[{\citenamefont{Gintis}(2000)}]{gintis_00}
\bibinfo{author}{\bibfnamefont{H.}~\bibnamefont{Gintis}},
  \emph{\bibinfo{title}{Game Theory Evolving}} (\bibinfo{publisher}{Princeton
  University Press}, \bibinfo{address}{Princeton}, \bibinfo{year}{2000}).

\bibitem[{\citenamefont{Nowak}(2006{\natexlab{a}})}]{nowak_06}
\bibinfo{author}{\bibfnamefont{M.~A.} \bibnamefont{Nowak}},
  \emph{\bibinfo{title}{Evolutionary Dynamics: Exploring the Equations of
  Life}} (\bibinfo{publisher}{Harvard University Press},
  \bibinfo{address}{Cambridge, MA}, \bibinfo{year}{2006}{\natexlab{a}}).

\bibitem[{\citenamefont{Nowak}(2006{\natexlab{b}})}]{nowak_s06}
\bibinfo{author}{\bibfnamefont{M.~A.} \bibnamefont{Nowak}},
  \bibinfo{journal}{Science} \textbf{\bibinfo{volume}{314}},
  \bibinfo{pages}{1560} (\bibinfo{year}{2006}{\natexlab{b}}).

\bibitem[{\citenamefont{Nowak and May}(1992)}]{nowak_n92b}
\bibinfo{author}{\bibfnamefont{M.~A.} \bibnamefont{Nowak}} \bibnamefont{and}
  \bibinfo{author}{\bibfnamefont{R.~M.} \bibnamefont{May}},
  \bibinfo{journal}{Nature} \textbf{\bibinfo{volume}{359}},
  \bibinfo{pages}{826} (\bibinfo{year}{1992}).

\bibitem[{\citenamefont{Lindgren and Nordahl}(1994)}]{lindgren_pd94}
\bibinfo{author}{\bibfnamefont{K.}~\bibnamefont{Lindgren}} \bibnamefont{and}
  \bibinfo{author}{\bibfnamefont{M.~G.} \bibnamefont{Nordahl}},
  \bibinfo{journal}{Physica D} \textbf{\bibinfo{volume}{75}},
  \bibinfo{pages}{292} (\bibinfo{year}{1994}).

\bibitem[{\citenamefont{Szab{\'o} and F{\'a}th}(2007)}]{szabo_pr07}
\bibinfo{author}{\bibfnamefont{G.}~\bibnamefont{Szab{\'o}}} \bibnamefont{and}
  \bibinfo{author}{\bibfnamefont{G.}~\bibnamefont{F{\'a}th}},
  \bibinfo{journal}{Phys. Rep.} \textbf{\bibinfo{volume}{446}},
  \bibinfo{pages}{97} (\bibinfo{year}{2007}).

\bibitem[{\citenamefont{Abramson and Kuperman}(2001)}]{abramson_pre01}
\bibinfo{author}{\bibfnamefont{G.}~\bibnamefont{Abramson}} \bibnamefont{and}
  \bibinfo{author}{\bibfnamefont{M.}~\bibnamefont{Kuperman}},
  \bibinfo{journal}{Phys. Rev. E} \textbf{\bibinfo{volume}{63}},
  \bibinfo{pages}{030901(R)} (\bibinfo{year}{2001}).

\bibitem[{\citenamefont{Ebel and Bornholdt}(2002)}]{ebel_pre02}
\bibinfo{author}{\bibfnamefont{H.}~\bibnamefont{Ebel}} \bibnamefont{and}
  \bibinfo{author}{\bibfnamefont{S.}~\bibnamefont{Bornholdt}},
  \bibinfo{journal}{Phys. Rev. E} \textbf{\bibinfo{volume}{66}},
  \bibinfo{pages}{056118} (\bibinfo{year}{2002}).

\bibitem[{\citenamefont{Holme et~al.}(2003)\citenamefont{Holme, Trusina, Kim, and Minnhagen}}]{holme_pre03}
\bibinfo{author}{\bibfnamefont{P.}~\bibnamefont{Holme}},
  \bibinfo{author}{\bibfnamefont{A.}~\bibnamefont{Trusina}},
  \bibinfo{author}{\bibfnamefont{B.~J.} \bibnamefont{Kim}}, \bibnamefont{and}
  \bibinfo{author}{\bibfnamefont{P.}~\bibnamefont{Minnhagen}},
  \bibinfo{journal}{Phys. Rev. E} \textbf{\bibinfo{volume}{68}},
  \bibinfo{pages}{030901} (\bibinfo{year}{2003}).

\bibitem[{\citenamefont{Masuda and Aihara}(2003)}]{masuda_pla03}
\bibinfo{author}{\bibfnamefont{N.}~\bibnamefont{Masuda}} \bibnamefont{and}
  \bibinfo{author}{\bibfnamefont{K.}~\bibnamefont{Aihara}},
  \bibinfo{journal}{Phys. Lett. A} \textbf{\bibinfo{volume}{313}},
  \bibinfo{pages}{55} (\bibinfo{year}{2003}).

\bibitem[{\citenamefont{Wu et~al.}(2005)\citenamefont{Wu, Xu, Chen, and Wang}}]{wu_pre05}
\bibinfo{author}{\bibfnamefont{Z.-X.} \bibnamefont{Wu}},
  \bibinfo{author}{\bibfnamefont{X.-J.} \bibnamefont{Xu}},
  \bibinfo{author}{\bibfnamefont{Y.}~\bibnamefont{Chen}}, \bibnamefont{and}
  \bibinfo{author}{\bibfnamefont{Y.-H.} \bibnamefont{Wang}},
  \bibinfo{journal}{Phys. Rev. E} \textbf{\bibinfo{volume}{71}},
  \bibinfo{pages}{037103} (\bibinfo{year}{2005}).

\bibitem[{\citenamefont{Lieberman et~al.}(2005)\citenamefont{Lieberman, Hauert, and Nowak}}]{lieberman_n05}
\bibinfo{author}{\bibfnamefont{E.}~\bibnamefont{Lieberman}},
  \bibinfo{author}{\bibfnamefont{C.}~\bibnamefont{Hauert}}, \bibnamefont{and}
  \bibinfo{author}{\bibfnamefont{M.~A.} \bibnamefont{Nowak}},
  \bibinfo{journal}{Nature} \textbf{\bibinfo{volume}{433}},
  \bibinfo{pages}{312} (\bibinfo{year}{2005}).

\bibitem[{\citenamefont{Tomassini et~al.}(2006)\citenamefont{Tomassini, Luthi, and Giacobini}}]{tomassini_pre06}
\bibinfo{author}{\bibfnamefont{M.}~\bibnamefont{Tomassini}},
  \bibinfo{author}{\bibfnamefont{L.}~\bibnamefont{Luthi}}, \bibnamefont{and}
  \bibinfo{author}{\bibfnamefont{M.}~\bibnamefont{Giacobini}},
  \bibinfo{journal}{Phys. Rev. E} \textbf{\bibinfo{volume}{73}},
  \bibinfo{pages}{016132} (\bibinfo{year}{2006}).

\bibitem[{\citenamefont{Vukov et~al.}(2008)\citenamefont{Vukov, Szab{\'o}, and Szolnoki}}]{vukov_pre08}
\bibinfo{author}{\bibfnamefont{J.}~\bibnamefont{Vukov}},
  \bibinfo{author}{\bibfnamefont{G.}~\bibnamefont{Szab{\'o}}},
  \bibnamefont{and} \bibinfo{author}{\bibfnamefont{A.}~\bibnamefont{Szolnoki}},
  \bibinfo{journal}{Phys. Rev. E} \textbf{\bibinfo{volume}{77}},
  \bibinfo{pages}{026109} (\bibinfo{year}{2008}).

\bibitem[{\citenamefont{Wang et~al.}(2006)\citenamefont{Wang, Ren, Chen, and Wang}}]{wang_pre06}
\bibinfo{author}{\bibfnamefont{W.-X.} \bibnamefont{Wang}},
  \bibinfo{author}{\bibfnamefont{J.}~\bibnamefont{Ren}},
  \bibinfo{author}{\bibfnamefont{G.}~\bibnamefont{Chen}}, \bibnamefont{and}
  \bibinfo{author}{\bibfnamefont{B.-H.} \bibnamefont{Wang}},
  \bibinfo{journal}{Phys. Rev. E} \textbf{\bibinfo{volume}{74}},
  \bibinfo{pages}{056113} (\bibinfo{year}{2006}).

\bibitem[{\citenamefont{Ren et~al.}(2007)\citenamefont{Ren, Wang, and Qi}}]{ren_pre07}
\bibinfo{author}{\bibfnamefont{J.}~\bibnamefont{Ren}},
  \bibinfo{author}{\bibfnamefont{W.-X.} \bibnamefont{Wang}}, \bibnamefont{and}
  \bibinfo{author}{\bibfnamefont{F.}~\bibnamefont{Qi}}, \bibinfo{journal}{Phys.
  Rev. E} \textbf{\bibinfo{volume}{75}}, \bibinfo{pages}{045101(R)}
  (\bibinfo{year}{2007}).

\bibitem[{\citenamefont{Chen and Wang}(2008)}]{chen_pre08}
\bibinfo{author}{\bibfnamefont{X.}~\bibnamefont{Chen}} \bibnamefont{and}
  \bibinfo{author}{\bibfnamefont{L.}~\bibnamefont{Wang}},
  \bibinfo{journal}{Phys. Rev. E} \textbf{\bibinfo{volume}{77}},
  \bibinfo{pages}{017103} (\bibinfo{year}{2008}).

\bibitem[{\citenamefont{Luthi et~al.}(2008)\citenamefont{Luthi, Pestelacci, and Tomassini}}]{luthi_pa08}
\bibinfo{author}{\bibfnamefont{L.}~\bibnamefont{Luthi}},
  \bibinfo{author}{\bibfnamefont{E.}~\bibnamefont{Pestelacci}},
  \bibnamefont{and}
  \bibinfo{author}{\bibfnamefont{M.}~\bibnamefont{Tomassini}},
  \bibinfo{journal}{Physica A} \textbf{\bibinfo{volume}{387}},
  \bibinfo{pages}{955} (\bibinfo{year}{2008}).

\bibitem[{\citenamefont{Santos and Pacheco}(2005)}]{santos_prl05}
\bibinfo{author}{\bibfnamefont{F.~C.} \bibnamefont{Santos}} \bibnamefont{and}
  \bibinfo{author}{\bibfnamefont{J.~M.} \bibnamefont{Pacheco}},
  \bibinfo{journal}{Phys. Rev. Lett.} \textbf{\bibinfo{volume}{95}},
  \bibinfo{pages}{098104} (\bibinfo{year}{2005}).

\bibitem[{\citenamefont{Santos et~al.}(2006)\citenamefont{Santos, Rodrigues, and Pacheco}}]{santos_prslb06}
\bibinfo{author}{\bibfnamefont{F.~C.} \bibnamefont{Santos}},
  \bibinfo{author}{\bibfnamefont{J.~F.} \bibnamefont{Rodrigues}},
  \bibnamefont{and} \bibinfo{author}{\bibfnamefont{J.~M.}
  \bibnamefont{Pacheco}}, \bibinfo{journal}{Proc. R. Soc. B}
  \textbf{\bibinfo{volume}{273}}, \bibinfo{pages}{51} (\bibinfo{year}{2006}).

\bibitem[{\citenamefont{Santos and Pacheco}(2006)}]{santos_jeb06}
\bibinfo{author}{\bibfnamefont{F.~C.} \bibnamefont{Santos}} \bibnamefont{and}
  \bibinfo{author}{\bibfnamefont{J.~M.} \bibnamefont{Pacheco}},
  \bibinfo{journal}{J. Evol. Biol.} \textbf{\bibinfo{volume}{19}},
  \bibinfo{pages}{726} (\bibinfo{year}{2006}).

\bibitem[{\citenamefont{G{\'o}mez-Garde{\~n}es et~al.}(2007)\citenamefont{G{\'o}mez-Garde{\~n}es, Campillo, Flor{\'{\i}}a, and Moreno}}]{gomez-gardenes_prl07}
\bibinfo{author}{\bibfnamefont{J.}~\bibnamefont{G{\'o}mez-Garde{\~n}es}},
  \bibinfo{author}{\bibfnamefont{M.}~\bibnamefont{Campillo}},
  \bibinfo{author}{\bibfnamefont{L.~M.} \bibnamefont{Flor{\'{\i}}a}},
  \bibnamefont{and} \bibinfo{author}{\bibfnamefont{Y.}~\bibnamefont{Moreno}},
  \bibinfo{journal}{Phys. Rev. Lett.} \textbf{\bibinfo{volume}{98}},
  \bibinfo{pages}{108103} (\bibinfo{year}{2007}).

\bibitem[{\citenamefont{Masuda}(2007)}]{masuda_prsb07}
\bibinfo{author}{\bibfnamefont{N.}~\bibnamefont{Masuda}},
  \bibinfo{journal}{Proc. R. Soc. B} \textbf{\bibinfo{volume}{274}},
  \bibinfo{pages}{1815} (\bibinfo{year}{2007}).

\bibitem[{\citenamefont{Poncela et~al.}(2007)\citenamefont{Poncela, G{\'o}mez-Garde{\~n}es, Flor{\'{\i}}a, and Moreno}}]{poncela_njp07}
\bibinfo{author}{\bibfnamefont{J.}~\bibnamefont{Poncela}},
  \bibinfo{author}{\bibfnamefont{J.}~\bibnamefont{G{\'o}mez-Garde{\~n}es}},
  \bibinfo{author}{\bibfnamefont{L.~M.} \bibnamefont{Flor{\'{\i}}a}},
  \bibnamefont{and} \bibinfo{author}{\bibfnamefont{Y.}~\bibnamefont{Moreno}},
  \bibinfo{journal}{New J. Phys.} \textbf{\bibinfo{volume}{9}},
  \bibinfo{pages}{184} (\bibinfo{year}{2007}).

\bibitem[{\citenamefont{Szolnoki et~al.}(2008)\citenamefont{Szolnoki, Perc, and Danku}}]{szolnoki_pa08}
\bibinfo{author}{\bibfnamefont{A.}~\bibnamefont{Szolnoki}},
  \bibinfo{author}{\bibfnamefont{M.}~\bibnamefont{Perc}}, \bibnamefont{and}
  \bibinfo{author}{\bibfnamefont{Z.}~\bibnamefont{Danku}},
  \bibinfo{journal}{Physica A} \textbf{\bibinfo{volume}{387}},
  \bibinfo{pages}{2075} (\bibinfo{year}{2008}).

\bibitem[{\citenamefont{Wu et~al.}(2006{\natexlab{a}})\citenamefont{Wu, Xu, Huang, Wang, and Wang}}]{wu_pre06}
\bibinfo{author}{\bibfnamefont{Z.-X.} \bibnamefont{Wu}},
  \bibinfo{author}{\bibfnamefont{X.-J.} \bibnamefont{Xu}},
  \bibinfo{author}{\bibfnamefont{Z.-G.} \bibnamefont{Huang}},
  \bibinfo{author}{\bibfnamefont{S.-J.} \bibnamefont{Wang}}, \bibnamefont{and}
  \bibinfo{author}{\bibfnamefont{Y.-H.} \bibnamefont{Wang}},
  \bibinfo{journal}{Phys. Rev. E} \textbf{\bibinfo{volume}{74}},
  \bibinfo{pages}{021107} (\bibinfo{year}{2006}{\natexlab{a}}).

\bibitem[{\citenamefont{Kim et~al.}(2002)\citenamefont{Kim, Trusina, Holme, Minnhagen, Chung, and Choi}}]{kim_pre02}
\bibinfo{author}{\bibfnamefont{B.~J.} \bibnamefont{Kim}},
  \bibinfo{author}{\bibfnamefont{A.}~\bibnamefont{Trusina}},
  \bibinfo{author}{\bibfnamefont{P.}~\bibnamefont{Holme}},
  \bibinfo{author}{\bibfnamefont{P.}~\bibnamefont{Minnhagen}},
  \bibinfo{author}{\bibfnamefont{J.~S.} \bibnamefont{Chung}}, \bibnamefont{and}
  \bibinfo{author}{\bibfnamefont{M.~Y.} \bibnamefont{Choi}},
  \bibinfo{journal}{Phys. Rev. E} \textbf{\bibinfo{volume}{66}},
  \bibinfo{pages}{021907} (\bibinfo{year}{2002}).

\bibitem[{\citenamefont{Szolnoki and Szab{\'o}}(2007)}]{szolnoki_epl07}
\bibinfo{author}{\bibfnamefont{A.}~\bibnamefont{Szolnoki}} \bibnamefont{and}
  \bibinfo{author}{\bibfnamefont{G.}~\bibnamefont{Szab{\'o}}},
  \bibinfo{journal}{Europhys. Lett.} \textbf{\bibinfo{volume}{77}},
  \bibinfo{pages}{30004} (\bibinfo{year}{2007}).

\bibitem[{\citenamefont{Wu et~al.}(2006{\natexlab{b}})\citenamefont{Wu, Xu, and Wang}}]{wu_cpl06}
\bibinfo{author}{\bibfnamefont{Z.-X.} \bibnamefont{Wu}},
  \bibinfo{author}{\bibfnamefont{X.-J.} \bibnamefont{Xu}}, \bibnamefont{and}
  \bibinfo{author}{\bibfnamefont{Y.-H.} \bibnamefont{Wang}},
  \bibinfo{journal}{Chin. Phys. Lett.} \textbf{\bibinfo{volume}{23}},
  \bibinfo{pages}{531} (\bibinfo{year}{2006}{\natexlab{b}}).

\bibitem[{\citenamefont{Perc and Szolnoki}(2008)}]{perc_pre08}
\bibinfo{author}{\bibfnamefont{M.}~\bibnamefont{Perc}} \bibnamefont{and}
  \bibinfo{author}{\bibfnamefont{A.}~\bibnamefont{Szolnoki}},
  \bibinfo{journal}{Phys. Rev. E} \textbf{\bibinfo{volume}{77}},
  \bibinfo{pages}{011904} (\bibinfo{year}{2008}).

\bibitem[{\citenamefont{Rong et~al.}(2007)\citenamefont{Rong, Li, and Wang}}]{rong_pre07}
\bibinfo{author}{\bibfnamefont{Z.}~\bibnamefont{Rong}},
  \bibinfo{author}{\bibfnamefont{X.}~\bibnamefont{Li}}, \bibnamefont{and}
  \bibinfo{author}{\bibfnamefont{X.}~\bibnamefont{Wang}},
  \bibinfo{journal}{Phys. Rev. E} \textbf{\bibinfo{volume}{76}},
  \bibinfo{pages}{027101} (\bibinfo{year}{2007}).

\bibitem[{\citenamefont{Flory}(1939)}]{flory_jacs39}
\bibinfo{author}{\bibfnamefont{P.~J.}~\bibnamefont{Flory}},
  \bibinfo{journal}{J. Am. Chem. Soc.} \textbf{\bibinfo{volume}{61}},
  \bibinfo{pages}{1518} (\bibinfo{year}{1939}).

\bibitem[{\citenamefont{Widom}(1966)}]{widow_jcp66}
\bibinfo{author}{\bibfnamefont{B.}~\bibnamefont{Widom}},
  \bibinfo{journal}{J. Chem. Phys.} \textbf{\bibinfo{volume}{44}},
  \bibinfo{pages}{3888} (\bibinfo{year}{1966}).

\bibitem[{\citenamefont{Dickman et~al.}(1991)\citenamefont{Dickman, Wang, and Jensen}}]{dickman_jcp91}
\bibinfo{author}{\bibfnamefont{R.}~\bibnamefont{Dickman}},
  \bibinfo{author}{\bibfnamefont{J.-S.}~\bibnamefont{Wang}}, \bibnamefont{and}
  \bibinfo{author}{\bibfnamefont{I.}~\bibnamefont{Jensen}},
  \bibinfo{journal}{J. Chem. Phys.} \textbf{\bibinfo{volume}{94}},
  \bibinfo{pages}{8252} (\bibinfo{year}{1991}).

\bibitem[{\citenamefont{Zimmermann et~al.}(2004)\citenamefont{Zimmermann, Egu{\'{\i}}luz, and San~Miguel}}]{zimmermann_pre04}
\bibinfo{author}{\bibfnamefont{M.~G.} \bibnamefont{Zimmermann}},
  \bibinfo{author}{\bibfnamefont{V.}~\bibnamefont{Egu{\'{\i}}luz}},
  \bibnamefont{and}
  \bibinfo{author}{\bibfnamefont{M.}~\bibnamefont{San~Miguel}},
  \bibinfo{journal}{Phys. Rev. E} \textbf{\bibinfo{volume}{69}},
  \bibinfo{pages}{065102(R)} (\bibinfo{year}{2004}).

\bibitem[{\citenamefont{Zimmermann and Egu{\'{\i}}luz}(2005)}]{zimmermann_pre05}
\bibinfo{author}{\bibfnamefont{M.~G.} \bibnamefont{Zimmermann}}
  \bibnamefont{and}
  \bibinfo{author}{\bibfnamefont{V.}~\bibnamefont{Egu{\'{\i}}luz}},
  \bibinfo{journal}{Phys. Rev. E} \textbf{\bibinfo{volume}{72}},
  \bibinfo{pages}{056118} (\bibinfo{year}{2005}).

\bibitem[{\citenamefont{Pacheco
  et~al.}(2006{\natexlab{a}})\citenamefont{Pacheco, Traulsen, and Nowak}}]{pacheco_jtb06}
\bibinfo{author}{\bibfnamefont{J.~M.} \bibnamefont{Pacheco}},
  \bibinfo{author}{\bibfnamefont{A.}~\bibnamefont{Traulsen}}, \bibnamefont{and}
  \bibinfo{author}{\bibfnamefont{M.~A.} \bibnamefont{Nowak}},
  \bibinfo{journal}{J. Theor. Biol.} \textbf{\bibinfo{volume}{243}},
  \bibinfo{pages}{437} (\bibinfo{year}{2006}{\natexlab{a}}).

\bibitem[{\citenamefont{Pacheco et~al.}(2006{\natexlab{b}})\citenamefont{Pacheco, Traulsen, and Nowak}}]{pacheco_prl06}
\bibinfo{author}{\bibfnamefont{J.~M.} \bibnamefont{Pacheco}},
  \bibinfo{author}{\bibfnamefont{A.}~\bibnamefont{Traulsen}}, \bibnamefont{and}
  \bibinfo{author}{\bibfnamefont{M.~A.} \bibnamefont{Nowak}},
  \bibinfo{journal}{Phys. Rev. Lett.} \textbf{\bibinfo{volume}{97}},
  \bibinfo{pages}{258103} (\bibinfo{year}{2006}{\natexlab{b}}).

\end{thebibliography}
\end{document}